\documentclass[aps,prb,twocolumn,10pt,superscriptaddress,citeautoscript,floats,
notitlepage,showpacs]{revtex4-1}

\usepackage{amsmath}
\usepackage{dcolumn}
\usepackage{graphicx}
\usepackage{xcolor}
\usepackage{bm}

\newcommand{\biox}{Bi$_2$O$_3$}

\begin{document}

\preprint{preprint}

\title{Two-dimensional electron gas in a metal/amorphous oxide interface with 
spin orbit interaction}

\author{Jose Manuel Flores-Camacho}
\email[e-mail (JMFC): ]{jmflores@cactus.iico.uaslp.mx}
\affiliation{Instituto de Investigaci\'on en Comunicaci\'on  \'Optica, 
Universidad Aut\'onoma de San Luis Potos\'{\i}, \'Alvaro Obreg\'on 64, 
78000 San Luis Potosi, Mexico}

\author{Jorge Puebla}
\email[e-mail (JP): ]{jorgeluis.pueblanunez@riken.jp}
\affiliation{Center for Emergent Matter Science, RIKEN, Wako, Saitama 
351-0198, Japan}

\author{Florent Auvray}
\affiliation{Center for Emergent Matter Science, RIKEN, Wako, Saitama 
351-0198, Japan}
\affiliation{Institute for Solid State Physics, University of Tokio, Kashiwa, 
Chiba 277-8581, Japan}

\author{Alfonso Lastras-Mart{\'\i}nez}
\affiliation{Instituto de Investigaci\'on en Comunicaci\'on  \'Optica, 
Universidad Aut\'onoma de San Luis Potos\'{\i}, \'Alvaro Obreg\'on 64, 
78000 San Luis Potosi, Mexico}

\author{Yoshichika Otani}
\affiliation{Center for Emergent Matter Science, RIKEN, Wako, Saitama 
351-0198, Japan}
\affiliation{Institute for Solid State Physics, 
University of Tokio, Kashiwa, Chiba 277-8581, Japan}

\author{Raul Eduardo Balderas-Navarro}
\affiliation{Instituto de Investigaci\'on en Comunicaci\'on  \'Optica, 
Universidad Aut\'onoma de San Luis Potos\'{\i}, \'Alvaro Obreg\'on 64, 
78000 San Luis Potosi, Mexico}

\date{\today}

\begin{abstract}
The  formation  of  novel  two-dimensional electron gas (2DEG) with 
high mobility in metal/amorphous interfaces 
has motivated an ongoing debate regarding the formation and novel 
characteristics of these 2DEGs. 
Here we report an optical study, based on infrared spectroscopic ellipsometry, of 
nonmagnetic metal and amorphous semiconducting 
oxide (Cu/Bi$_2$O$_3$) interfaces that confirms the formation of a 2DEG 
with spin orbit coupling (SOC). 
The 2DEG optical response was simulated with a uniaxial diagonal dielectric tensor
within a sub-nanometer thin layer, where
its $x$ and $z$ components lineshapes resolved in both free-electron and 
peak-like contributions, resulted very similar to theoretical predictions 
[M. Xie \emph{et al.}, Phys. Rev. B {\bf 89}, 245417 (2014)] 
of a two dimensional electron gas confined in the normal direction 
of a perovskite interface. In particular, the small but finite conducting character 
of the 
$z$ component provides a unambiguous signature of the presence of the 2DEG
in the Cu/\biox\ system.
Although the original constituent materials do not possess 
spin-orbit coupling (SOC), the resulting interfacial 
hybridization of such states induce electronic asymmetric wave functions. 
This work demonstrates the detection of 2DEG in amorphous crystals 
allowing to study its challenging interfacial phenomena such as SOC and 
interface-bulk coupling, overcoming an experimental impediment which has 
hold back for decades important advancements for the understanding 
of 2DEGs in amorphous materials.
\end{abstract}


\maketitle

\section{Introduction}

The concept of two-dimensional electron gas (2DEG) has 
contributed enormously 
to the understanding of the rich phenomena of electrons 
confined at surfaces 
and interfaces. In high quality semiconductor 
heterostructures, 2DEG 
confinement induced enhancement of electron mobility leading to 
technology advancements and scientific discoveries, such as the 
fractional Quantum Hall 
effect~\cite{Horst1999}. The 
continuous development of growth techniques allows nowadays to 
obtain enough 
good quality heterostructures from oxides, forming novel 2DEGs. 
Interestingly, 2DEG in oxides show a significant increase in 
electron 
density and strong electron correlation, with consequences in 
magnetic properties, superconductivity, ferroelectricity and 
spin orbit 
interaction~\cite{Mannhart2010, Tokura2012}. Intrinsically, the 
electronic structure 
of 2DEG in oxide interfaces is different from the most standard 
semiconductor counterparts. In most of the oxide interfaces, 
the transport properties are dominated by narrow $d$-band 
electrons, 
whereas in semiconductors the electrons localized at states 
at the bottom of the conduction band dictate the properties of 
the 2DEGs. 
However, despite of the great advance in the understanding of 
2DEG in complex oxide interfaces, still many oxide interfaces 
are challenging to characterize, mainly because their crystal 
quality. 
Analysis of the electronic structure is 
commonly achieved by angle-resolved photoemission spectroscopy 
(ARPES). 
However, ARPES characterization requires high quality, 
large and flat 
crystals. For instance, high quality 
crystals of bismuthates were obtained, allowing detailed 
characterization by ARPES and reveling the mechanism of its 
high temperature superconductivity, an open discussion for more 
than 30 years~\cite{Wen2018}. 
 
Moreover, the formation of novel 2DEG with high mobility in 
amorphous oxides 
has motivated an increasing 
interest~\cite{Woon2012, Chen2015}. The origin of the 2DEG 
formation at these 
interfaces has been assigned to electronic reconstruction via 
interfacial charge transfer~\cite{Liu2013}; however, 
detailed characterization of the properties of 2DEGs formed by 
amorphous oxides is still lacking. An 
alternative to ARPES characterization is to analyze the optical 
conductivity of the 2DEGs 
by polarized spectroscopy techniques such 
as spectroscopic ellipsometry. Although, structural distortion 
and defects in amorphous materials may cause complex 
dielectric screening, and significantly complicate the 
analysis of the properties of the 2DEG, a theoretical model 
can be assessed and improved by direct feedback from 
optical characterization~\cite{xie14prb}. 

Here, we report on mid-infrared ellipsometry characterization of 
the properties of 2DEG formed at the interface between a 
nonmagnetic metal and amorphous 
semiconducting oxide, Cu/Bi$_2$O$_3$. Recent spin-charge 
interconversion experiments suggest the presence of a 2DEG 
with spin orbit coupling (SOC) at 
this interface~\cite{puebla17apl, Xu2018, tsai18sciRep}, 
making it a very attractive structure for spin based 
complementary metal-oxide semiconductor 
technologies (CMOS)~\cite{Young2018,Manipatruni18nat}. The 
characterization by angle resolved ellipsometry confirms the 
formation of a 2DEG with SOC. In-depth analysis of 
the optical conductivity, modeled by a realistic electronic 
structure system and Kubo formalism, allows to define the 
origin of SOC as the hybridization of interfacial Cu-O-Bi 
states by charge transfer, generating an asymmetric wave 
function~\cite{tsai18sciRep}, as 
previously suggested for amorphous perovskite 
oxides~\cite{Chen2015, Liu2013}. 

The rest of the paper is organized as follows. The main and reference samples are described in Sec.~\ref{secc:exp} together
with a brief description of the ellipsometry technique. In Sec.~\ref{sec:model}
we present the dielectric functions of the stacked system and the
model for the tensorial dielectric function of the 2DEG forming at the
Cu/\biox\ interface. Finally, results and discussion are given in Sec.~\ref{sec:results}.

\section{Experiment}
\label{secc:exp}

\begin{figure}[htbp]
 \includegraphics[width = 8.0 cm]{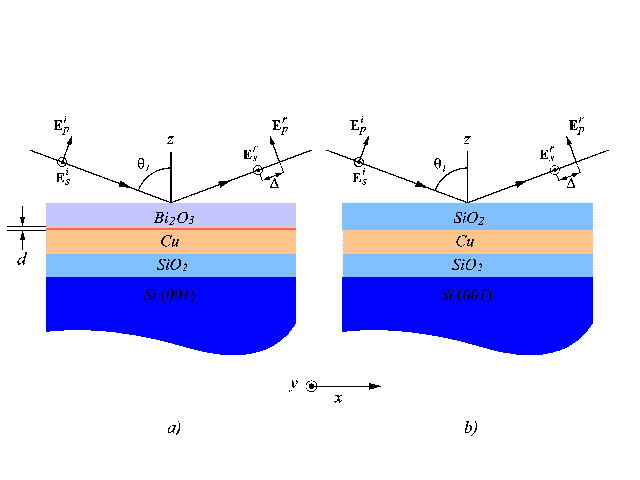}
    \caption{(Color online) Diagrams of the samples used for 
    the ellipsometry measurements at angles of 
    incidence $\theta_i$. (a) Main samples 
    Si(001)/SiO$_2$/Cu/\biox\
    (Cu thicknesses are 10, 20, or 50~nm). The 2DEG 
    formed upon deposition of the Bi$_2$O$_3$ film is 
    depicted with thickness $d$. (b)
    reference sample with SiO$_2$ instead of 
    Bi$_2$O$_3$. $x$, $y$ and $z$ are the axes for which the tensorial complex components $\varepsilon_{xx} = \varepsilon_{yy}$ and 
    $\varepsilon_{zz}$ of the 2DEG are related. $E_p$ and $E_s$ stand for $p$ and $s$ polarizations for 
    both incident and reflected optical electric fields, and $\Delta$ is a measure for the retardance 
    between $p$ and $s$ polarizations upon reflection. See text for details.}
    \label{fig:Scheme}
\end{figure}

The samples consist of Bi$_2$O$_3$ films, with thickness of 
20~nm, grown on 
a previously (either 10, 20, or 50~nm thick) 
Cu-capped SiO$_2$/Si(001) substrate
as described elsewhere~\cite{puebla17apl, florent2018}.  
The structure is shown in  Fig.~\ref{fig:Scheme}(a).
Additionally, in order to contrast the presence of 2DEG and
for obtaining the dielectric function (DF) of thin Cu film 
alone, a second set of reference samples were prepared as 
shown in Fig.~\ref{fig:Scheme}(b). Two other samples,
Si/SiO$_2$ and Si/SiO$_2$/Bi$_2$O$_3$ were 
made in order to measure the SiO$_2$ film thickness of 
all samples' common substrate, and to determine the
dielectric function of Bi$_2$O$_3$ in the present spectral range.

The samples were characterized by infrared spectroscopic 
ellipsometry (IRSE) 
by means of an IR-VASE apparatus 
(J.A. Woollam Co.). Ellipsometry measures the change in polarization state upon
oblique reflection of an originally linear 
polarized incident 
beam~\cite{Azzam77}.
The reflected state of polarization is measured by the 
complex ratio
\begin{equation}
\label{eq:psi-del}
r_p/r_s = \tan\psi\,\exp(i \Delta),
\end{equation}  
\noindent where $r_p$ and $r_s$ are the complex 
reflection coefficients 
for $p$ and $s$ polarizations of light,
standing for parallel and perpendicular to the 
plane of incidence, respectively. 
The ellipsometric angles $\psi$ and 
$\Delta$ measure then the relative change of amplitude 
and phase, respectively, of the $p$ to $s$ 
polarizations~\cite{humlicek05tompkins}.
Mid-infrared SE (spectroscopic ellipsometry) was measured in the range of $\sim$35 to 
760~meV in ambient conditions for angles of incidence
$\theta_i = 35^\circ, 45^\circ, ..., 85^\circ$. As the indicatrix of an
uniaxial material is oriented with its distinct axis perpendicular to the sample surface,
i.e., $\varepsilon_1=\varepsilon_2\neq \varepsilon_3 (= \varepsilon_z)$ an ordinary
so-called isotropic-like ellipsometry measurement is sufficient to fully characterize the
optical properties~\cite{FujiwaraSEllipsom}. Neither off-diagonal Jones matrix 
elements nor M\"uller matrix measurements provided further information for our samples.

\begin{figure}[htbp]
 \includegraphics[width = 7.5 cm]{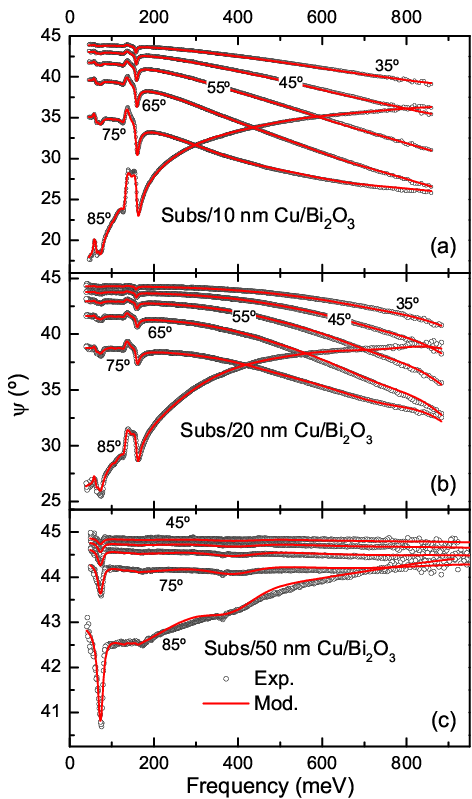}
    \caption{
    (Color online) Experimental (symbols) and 
    calculated (lines) $\psi$ spectra recorded at 
    different angles of incidence (indicated) corresponding 
    to Bi$_2$O$_3$ on (a) 10~nm Cu, (b) 20~nm, and 
    (c) 50~nm Cu layers, respectively. 
    }
    \label{fig:fitExp}
\end{figure}

For the mid-infrared studies, where silicon is transparent, partially incoherent light
reflected off the backside surface can reach the detector. Thus the
backside of the samples were subjected to a sand-blasting treatment,
obtaining thus a surface rough enough so as to avoid specular, spurious 
reflection from it. The success of the treatment is tested by transmission 
and depolarization measurements, which are acquired 
with the same ellipsometer. Transmission turned out to be negligible
for all samples, whereas the loss of polarization yielded near zero values as well,
except for samples with thicker Cu film and only for $\theta_i=85^\circ$, 
for which the loss is below 2.5\%. This may indicate another source of depolarization.
Resulting experimental spectra are shown with symbols in Fig.~\ref{fig:fitExp}. 

\section{Ellipsometry Model}
\label{sec:model}

In this section we describe the procedure to obtain the 
model employed in the present work for the diagonal, uniaxial, dielectric tensor 
representing the optical response of the 2DEG.
We present first the model of stacked layers and the dielectric
functions of the corresponding constituent materials followed 
by the 2DEG dielectric tensor.
The model the 2DEG layer is stripped to the minimum number of oscillators that
permitted a good fit. We note that the core result of this work, shown in 
Fig.~\ref{fig:fitExp},
presents some deviations from experiment which might have been corrected by including 
more oscillators, however, this is avoided if a physical interpretation cannot 
be provided.  

The layer basic constituents' dielectric functions, i.e., those of Si, SiO$_2$, 
the thickness dependent Cu, and \biox, were obtained either from literature or extracted 
from reference samples, so 
that for the main Cu/\biox\ samples, the only \emph{permitted} unknown is the 
2DEG response. Moreover, spectra corresponding to all angles of incidence are fitted 
together employing the same constituents. This ensures consistency of the model.  

\subsection{Constituent materials}

\begin{figure}[htbp]
  \includegraphics[width = 7 cm]{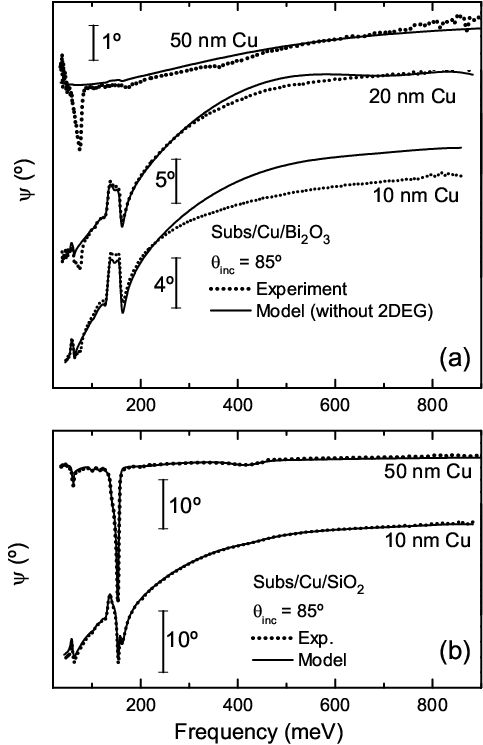}
    \caption{
    Calculated (solid lines) and experimental (dashed lines) $\psi$ spectra at an angle of incidence of 
    85$^\circ$ corresponding to 
    (a) the main samples, Subs./Cu/\biox\ for the indicated Cu thicknesses, for which a 2DEG interfacial layer is not included.
    These are to be considered the starting point of the modeling.
    (b) Spectra of  
    Subs./Cu/SiO$_2$-type reference samples. This kind of samples are used to 
    determine the Cu thickness-dependent
    optical conductivities presented in Fig.~\ref{fig:Constituents}(c). 
    In both, the vertical scales correspond to units of ellipsometric angle $\psi$:
    each spectrum is scaled differently to aid visualization.
    }
\label{fig:sio2-psi}
\end{figure}

The total reflection coefficients~\cite{osheavens} in Eq.~(\ref{eq:psi-del}) are calculated 
separately for $s$ and $p$ polarizations using Fresnel 
coefficients consisting of, for the main Cu/\biox\ samples, 
a model of 6 stacked media: 
(0) vacuum, (1) 20~nm Bi$_2$O$_3$, (2) an 
interfacial, anisotropic layer to accommodate the 2DEG, 
(3) Cu thin film, (4) 320~nm SiO$_2$, and
(5) Si substrate [see Fig.~\ref{fig:Scheme}(a)]. 
Similar models are used for the different reference samples,
for instance, Fig.~\ref{fig:Scheme}(b) corresponds to
Si/SiO$_2$/Cu/SiO$_2$.
It is important to note that attempts made to model the Cu/\biox\  spectra without 
the artificially introduced 2DEG layer failed to simulate 
the main features of the 
experimental curves as presented in Fig.~\ref{fig:sio2-psi}(a). However, it worked nicely for the 
Cu/SiO$_2$ reference samples, as shown in 
Fig.~\ref{fig:sio2-psi}(b), indicating that the 
interfacial layer must be necessarily included in the
description of the optical response of the Cu/Bi$_2$O$_3$ system.

In order to ensure Kramers-Kronig consistency, 
the complex dielectric function of each layer is parameterized as 
a sum of oscillators of a general form
\begin{equation}
\label{eq:gen:DF}
\varepsilon(\omega) = \varepsilon_\infty + 
\varepsilon_{\mathrm{D}}(\omega)
+ \varepsilon_{TL}(\omega) +\sum_i \varepsilon_{L,i}(\omega),
\end{equation}
\noindent where $\varepsilon_\infty$ is a (real) constant offset, the 
$\varepsilon_{L,i}(\omega)$ terms 
are used to include resonances within the measured spectral range, commonly Lorentz
or complex Gaussian functions. $\varepsilon_{TL}$ 
is the Tauc-Lorentz oscillator~\cite{jellison05tompkins},
which is used here because of its pulse-like shape as described below.
The $\varepsilon_{\mathrm{D}}(\omega)$ term corresponds to the classical  
Drude model for free carriers expressed through 
\begin{equation}
 \varepsilon_\mathrm{D}(\omega) = -\frac{\sigma^0}{\varepsilon_0 \omega\left(i+\tau\omega\right)},
\end{equation}
\noindent where the offset with which this equation is commonly written is passed 
to $\varepsilon_\infty$ in Eq.~(\ref{eq:gen:DF}), and $\sigma^0$ is the DC conductivity, which, for the case of the interfacial 2DEG, 
will be replaced by the appropriate tensorial component $\sigma_{ij}^0$.
Similarly, in this layer the characteristic life time $\tau$ will be direction dependent, as well.
However, if the material of the layer is around the 
percolation threshold, the Drude term
can be replaced by a Drude-Smith line shape given by~\cite{smith01prb} 
\begin{equation}
    \varepsilon_\mathrm{DS}(\omega)  =  \frac{i}{\varepsilon_0 \omega}\frac{\sigma^0}{1-i\omega\tau}\left(1+\frac{c}{1-i\omega\tau}\right)
\end{equation}
which accounts for stochastic 
persistency of velocity of carriers after scattering events
measured through the factor $c$ ($-1<c<0$).
In this model $\tau$ will be the average collision time, and thus the real DC conductivity, i.e., the one obtained
by extrapolation to $\omega\rightarrow 0$ is reduced with respect to that of the pristine material.
The Drude-Smith model predicts a shift of spectral weight which is expressed by a broad 
peak in the mid-infrared region~\cite{smith01prb} in agreement with
classical percolation theories~\cite{stroud79prb}. The required Lorentzian and Tauc-Lorentz oscillators
are only phenomenological descriptions for the physical reality of the spin-orbit coupling~\cite{xie14prb} related bands and inter-subband transitions that
serve to place experimentally those resonances.

\begin{figure}[htbp]
 \includegraphics[width = 7.0 cm]{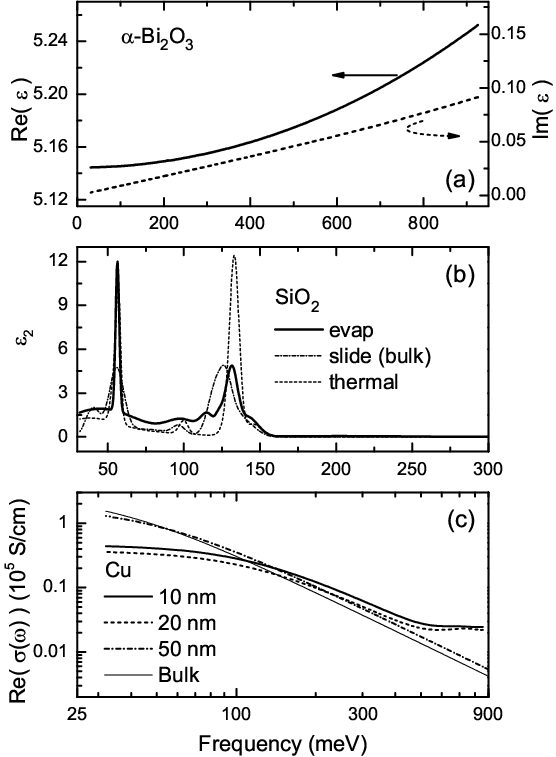}
    \caption{Dielectric functions (or optical conductivities) of the basic constituents of the optical model.
    (a) Complex dielectric function of $\alpha$-Bi$_2$O$_3$ as obtained from a reference sample. 
    (b) $\varepsilon_2$ of SiO$_2$. The labels \emph{evap} and \emph{thermal} correspond to the evaporated
    layer for reference samples with no 2DEG formation, and the substrate for all samples in the present work, respectively.
    \emph{slide (bulk)}, is presented for comparison. Notice the shorter spectral range used to
    emphasize the region of SiO$_2$ characteristic vibrations. The SiO$_2$ reference samples 
    (see Fig.~\ref{fig:sio2-psi}) were employed to
    extract the Cu thickness-dependent optical conductivities depicted in (c). 
    Cu bulk data were taken from Ref.~[\onlinecite{palik1985}] for guidance.}
\label{fig:Constituents}
\end{figure}

The dielectric function of $\alpha$-Bi$_2$O$_3$ was extracted from a 
Si/thermal SiO$_2$/\biox\ reference sample.
To this end a reference substrate Si/SiO$_2$ was fitted first. The thermal SiO$_2$ dielectric 
function was taken from data sets without modifications.
It is shown in Fig.~\ref{fig:Constituents}(b) (dashed line).
The resulting complex dielectric function $\varepsilon$ of \biox\ is shown in 
Fig.~\ref{fig:Constituents}(a). 

Employment of Si/SiO$_2$/Cu/SiO$_2$ reference samples was required to extract the thickness 
dependent Cu dielectric function, since in this samples no 2DEG formation is expected. Due to the preparation method,
i.e., evaporation instead of thermal formation, the dielectric function of the most external 
SiO$_2$ layer did require some modification. Its $\varepsilon$ resembles more that of a bulk glass (a microscope slide). In 
Fig.~\ref{fig:Constituents}(b) we show the spectral region where the preparation-dependent differences are stronger~\cite{kitamura07ao}. 

For the present Cu thin layers, the bulk dielectric function cannot be used since for these thicknesses the 
Cu layers are still near the mentioned percolation threshold~\cite{hoevel10prb,lee11jjap,Amothchkina11ao} even though they are already continuous.
The layers of 20 and 50~nm required the classical Drude model only, however, for the 10~nm Cu film the
Drude-Smith correction was necessary to better fit the Cu/SiO$_2$ data of Fig.~\ref{fig:sio2-psi}(a),
since the model without the correction had the tendency of rising above the experimental data at the lower end of the spectrum. 
The resulting Cu optical conductivities are shown in Fig.~\ref{fig:Constituents}(c), where bulk data~\cite{palik1985} are shown 
for comparison.

\subsection{Anisotropic model for the 2DEG}

The 2DEG layer has drastically different transport properties along the Cu/\biox\ interface and perpendicular to it.
This strong disparity prompted us to pursue a simulation-to-experiment approach instead of the common experiment-to-theory procedure.
To extract its frequency dependent tensorial complex components $\varepsilon_{xx} = \varepsilon_{yy}$ and 
$\varepsilon_{zz}$, we first proceeded with the assumptions that the conductivity along the interface is much greater 
than that of the out-of-plane component, and that a forced isotropic approximation could suffice to provide
a preliminar lineshape as a first approximation, and thus, learn already something about the main spectral features as described next.
A similar anisotropic approach has been conducted for perovskites characterized also by infrared ellipsometry~\cite{yazdi16epl} in which a Berreman resonance 
is excited by the presence of a longitudinal phonon in one of their constituents. In our case, on the other hand, the properties of the
out-of-plane conductivity, in particular a sort of ``bulk'' plasmon-like characteristic frequency, and not a longitudinal phonon,  
are critical for the description of the present experiments as discussed in the following.

The extracted ``isotropic'' optical properties are to be considered a pseudo dielectric function $\langle\varepsilon\rangle$ of the layer: 
from a set of pre-simulations we first inferred that $\langle\varepsilon\rangle \neq \varepsilon_{ij}\ell_i\ell_j$ 
(the modulus of a tensor along a given direction) even for bulk-like transparent well-oriented uniaxial media, where $i,j=(x,y,z)$ and $\ell_i$ are direction cosines
of the traveling electric displacement vector $\mathbf{D}$. This implies that, as we learned from simulations,
in the case of free electron behavior described by Drude line shape of the form $(\rho_0,\tau)$,
where $\rho_0$ is the dc resistivity and $\tau$ is the characteristic time of life,
that when the dc parts of the optical conductivity moduli follow the relation $\sigma^0_{xx} > \sigma^0_{zz}$
the isotropically forced conductivity $\langle\sigma\rangle$ tends to overestimate the dc $\sigma_x$ component even by an order of magnitude.
However, the isotropic-like time of life $\tau_\mathrm{iso}$ turns out to be $\tau_x$ along the interface, and tends to be always 
correctly predicted at least under the aforementioned dc conductivities relation. 
Interestingly, the bulk plasmon frequency $\omega_p$; i.e., the crossing
Re$\, \varepsilon(\omega_p)=0$, produces a resonant-like feature in ellipsometry spectra where there is no actual resonance. 
In fact, for the isotropic forced simulation, this peak was simulated with a Fano resonance due to its asymmetric 
form~\cite{fano61pr,lukyanchuk10natMat}, which can be seen more clearly in the experimental spectrum at $\theta_i = 85^\circ$ in Fig.~\ref{fig:fitExp}(c), 
and was even seriously considered as a candidate to be a contribution of either $\varepsilon_{xx}$ or $\varepsilon_{zz}$ were not for the simpler 
explanation provided by the  out-of-plane conductivity.
Moreover, attempts to introduce a Fano resonance resulted in non physically negative dielectric functions. 
Therefore, any simulation taking into account Fano resonances was excluded in the anisotropic model.
The asymmetric negative peak was thus obtained in simulations as follows: 
the feature is quite pronounced in calculated spectra of bulk-like samples if any one of $\varepsilon_x$ and/or $\varepsilon_z$ 
has this crossing, 
but for films on Cu substrates this apparent resonant feature is only revealed in calculated ellipsometry spectra 
for the $\varepsilon_z$ crossing and not for $\varepsilon_x$ even if this last function has it. We can therefore regard the presence
of this feature, as the one at around 70 - 80~meV in Fig.~\ref{fig:fitExp}, as a signature of the formation of an anisotropic layer
at the Cu/\biox\ interface, which in our case is extremely revealing since our samples are composed of amorphous materials and 
therefore have isotropic dielectric functions. 
To sustain this proposition we note that ($i$) this is not a SiO$_2$
vibration: the frequencies do not exactly correspond and for the sample with thicker Cu layer, as seen in Fig.~\ref{fig:fitExp}(c), 
Cu clearly quenches any SiO$_2$ feature, but the $\varepsilon_z$ crossing related peak is very dominant. ($ii$) This feature
should not be confused with a Berreman peak~\cite{berreman63pr} or other similar peak-like features, since those occur either close to 
actual resonances~\cite{berreman63pr}, when the substrate has a crossing~\cite{nucara18prb}  Re$\,\varepsilon=1$ with positive slope
(and not Re$\,\varepsilon=0$, as in our case),
or when the real part of the dielectric functions at both sides of the interface
coincide~\cite{park13prb,dubroka10prl}, respectively. We clearly have not the conditions to 
fulfill any of these criteria.

For Lorentz or Gaussian line shapes, as the ones assigned to spin-orbit coupling bands in the next section, 
simulations showed that both characteristic centers and broadenings can be already correctly inferred from
isotropic-like simulations. The real parts (Re$\,\varepsilon_{L}$)$_x$ and (Re$\,\varepsilon_{L}$)$_z$, which describe the elastic response of the media, 
are inverted with respect to each other. For layered media, the real and imaginary parts seem to be interchanged in the total
ellipsometry spectra. All these criteria provide guidance to decide whether a feature is to be part of the $x$ or $z$ 
component of the $\varepsilon$ tensor, based also on reduction of fitting error. 

Concerning film thicknesses, the forced isotropic ``best'' fit also overestimates film thicknesses of the supposed 2DEG layer.
The obtained artificial layer thickness forming between Cu and \biox\ resulted in 2 to 3.6~nm depending on the underlying Cu film thickness
for the forced-isotropic model, whereas for the more realistic uniaxial model, the thicknesses are all between 0.5 and $\leq$1~nm
(see Table.~\ref{tab:param}, below).  

Finally, for the 2DEG optical properties we propose the following description: 
the two-dimensional character of the electron gas
will be described through a very thin layer with dielectric function of 
tensorial, uniaxial character, for which the
carrier transport is much easier along the interface. 
The in-plane tensor modulus $\varepsilon_{xx}$ will have a Drude 
(or Drude-Smith) component to accommodate the free electron contribution 
to the 2DEG, and two Gaussian oscillators to simulate the 
SOC contributions~\cite{xie14prb}. The out-of-plane $\varepsilon_{zz}$ 
modulus will be comprised of a free electron Drude 
contribution, for which the relation $\sigma^0_{zz}\ll\sigma^0_{xx}$ holds, 
but which is nevertheless finite in such a way that it has 
a ``bulk'' plasmon Re$\,\varepsilon_{zz}(\omega_{p,z})=0$ crossing,
which is critical in our model.
Additionally, $\varepsilon_{zz}$ includes a Tauc-Lorentz oscillator whose role,
expected by theoretical considerations~\cite{xie14prb}, 
will be described in the next section. 
The model as such is simple as desired, and quite descriptive of 
the actual experiments as seen in Fig.~\ref{fig:fitExp}.
Each spectral feature
is modeled with two (for the classical Drude) to four parameters (for the percolation modified Drude-Smith model),
which apparently amount to a large number of free quantities to determined, however, most of the spectral features are well
separated from the rest and can be thus considered as independent from the others. Furthermore,
the isotropic-forced first approach already provides a good estimation of the values some of them, specifically,
the in-plane free-electron life time and positions of both SOC peaks.
Additionally, the correlation between some parameters, and its possible source, is discussed at the end of the main text.

\section{Results and discussion}
\label{sec:results}

The substrate Si/SiO$_2$, and the Si/SiO$_2$/Bi$_2$O$_3$ 
reference samples were fitted with VASE32, obtaining thus 
the thickness of the thermally grown SiO$_2$ of 322~nm,
 and the mid-infrared 
\emph{pristine} Bi$_2$O$_3$ dielectric function of Fig.~\ref{fig:Constituents}(a),
which are used for all subsequent samples. 
$\alpha$-Bi$_2$O$_3$ has 
a series of small phonons in this spectral region below 
66~meV~\cite{narang94jms,kuzmenko96jltp}, which were 
not detected in our experiment. Also, the group of features 
at $\sim\!60$ and $150$~meV in Fig.~\ref{fig:fitExp}(a), 
(b), and Fig.~\ref{fig:sio2-psi}(a),(b)
correspond to SiO$_2$ 
and their strengths, when SiO$_2$ is used only as the substrate of Cu/\biox, 
can therefore be employed as a gauge of spectral absorption by the different Cu layers. 
For the SiO$_2$/Cu/SiO$_2$ reference spectra, shown in
Fig.~\ref{fig:sio2-psi}, the dips are
easily reproduced by the proper model without great effort.

\begin{figure}[htbp]
 \includegraphics[width = 7.5 cm]{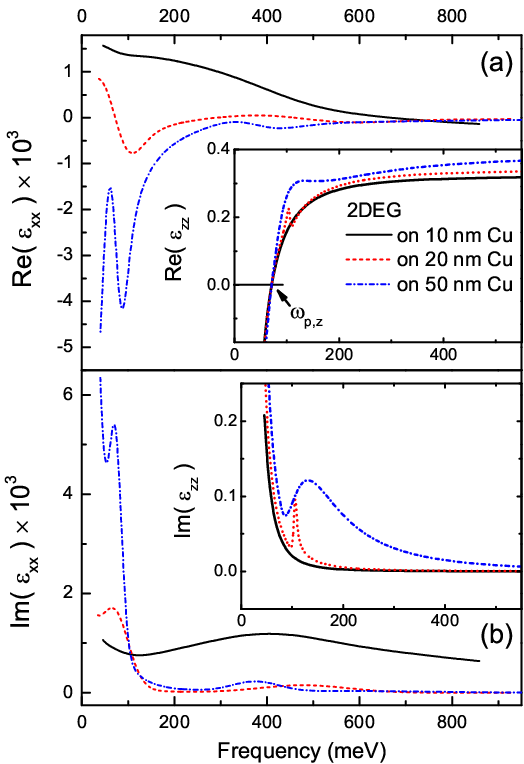}
    \caption{(Color online)
    Real (a) and imaginary (b) parts of the diagonal dielectric tensor components of 
    the 2DEG interfacial layer forming between Bi$_2$O$_3$ and Cu as extracted from the fitting procedure for the 
    samples with Cu layer thickness of 10, 20, and 50~nm, as indicated. The main plots show $\varepsilon_{xx}$
    and the insets correspond to $\varepsilon_{zz}$. The plasmon frequency $\omega_{p,z}$ is indicated 
    with a horizontal line at the crossing Re$(\varepsilon_{zz}) = 0$.}
\label{fig:sigmaAll}
\end{figure}

Experimental and calculated $\psi$ spectra for 
the different Cu/Bi$_2$O$_3$ samples recorded at 
several angles of incidence are shown in 
Fig.~\ref{fig:fitExp}(a)-(c). 
The overall spectra are dominated by Cu reflectivity and 
its corresponding thickness-dependent optical conductivity
$\sigma(\omega) = i \omega \varepsilon_0 \left(1-\varepsilon(\omega)\right)$. 
 The resulting tensorial dielectric complex moduli, $\varepsilon_{xx}$ and
 $\varepsilon_{zz}$  
of the 2DEG forming at the Cu/Bi$_2$O$_3$ 
interface are presented in Fig.~\ref{fig:sigmaAll}. 
The resulting dielectric properties depend on the underlying Cu film 
thickness-driven transport properties. This dependence can be 
explained in terms of modulation of charge transfer from Cu
to Bi$_2$O$_3$ due 
to the mobility of electrons within the Cu layer itself: 
for thicknesses below 20~nm  
the layer is below its percolation threshold (in fact,
10~nm Cu was modeled with the Drude-Smith approach),
which implies the formation of potential barriers within the Cu
layer; very likely produced by plane defects at the domain connections, and consequently modified work functions. 
Indeed, it has been reported that 
the relation of work functions of the materials 
at the interface is critical for its resulting
characteristics, like the relative
occupancy of hybridized orbitals near the
interface~\cite{park13prb}, and thus, those of the 2DEG
at the Bi$_2$O$_3$ on non-magnetic metals 
interfaces~\cite{tsai18sciRep} and perovskite
oxide-oxide interfaces, as well~\cite{Chen2015}.

The new layer, in contrast to the aforementioned Cu/SiO$_2$ experiments, 
resulted in strong optical spectral features 
quite different from those of the original constituent materials
as evinced by comparing Fig.~\ref{fig:sigmaAll} and Figs.~\ref{fig:Constituents}(a) and (c). 
Strikingly, the line shape of the interfacial layer 
DF thus resolved in Drude and other resonant oscillator 
components, resulted very similar to theoretical 
predictions~\cite{xie14prb}
of a two dimensional electron gas confined in the normal 
direction of 
the system. The hybridization of interfacial orbital states 
induce electronic asymmetric wave functions
and splitting of spin-orbitals, in contrast with
the original constituent materials which do not possess 
SOC. Such asymmetric wave functions
thus induce an interfacial electric field suggesting
SOC of the Rashba type~\cite{bychkov84pzetf,manchon15natMat}. 
This statement is in agreement with first-principle calculations recently 
reported for Cu/Bi$_2$O$_3$~\cite{tsai18sciRep, JPuebla2019}.

As mentioned above, the 2DEG $\varepsilon_{xx}$ and $\varepsilon_{zz}$ moduli are very different.
The scales in Fig.~\ref{fig:sigmaAll} differ by four orders of magnitude between $x$ and $z$ 
components
and the spectral structures are very dissimilar.
Had we kept an isotropic approach for all layers, the fitting error would grow with increasing 
angle of incidence. This error, although present, is misleadingly small and could have been easily considered
as random instead of systematic.
The 2DEG anisotropic dielectric functions in Fig.~\ref{fig:sigmaAll} are 
dominated by
the free electron contribution, which is modeled by a classic Drude line shape for 
the 20 and 50~nm Cu samples.
The 2DEG free electron part corresponding to the 10~nm Cu sample was simulated with the 
Drude-Smith approach~\cite{smith01prb}. 
The broad spectral structure centered at $\sim\!$400~meV is then a sign of abundance of scattering 
events, and is consistent then, with phenomenological percolation theories producing 
a mid-infrared broad peak accompanying the main
Drude feature, and predict a recovery of the usual 
free-electron gas behavior as the percolation threshold 
is surpassed~\cite{smith01prb}. For perovskites systems
this metal-insulator transition has been studied as a function of temperature and by 
modulating carrier densities by means of gating~\cite{ahadi17prl,ahadi17apl}.
Here, the carrier density and the Rashba effect-inducing interfacial electric field are provided by the
dielectric characteristics of the underlying Cu supporting film. 
The parameters,
such as dc conductivities $\sigma^0_{xx}$, $\sigma^0_{zz}$, and mean times of life $\tau_{xx}$ 
and $\tau_{zz}$ of the
free electron contribution to the 2DEG are presented in Table~\ref{tab:param}
as functions of underlying Cu film thickness.
It is observed that while 
$\sigma^0_{xx}$ increases non monotonically, but fast with increasing Cu 
thickness, it does not show a clear one-to-one relation with the corresponding Cu-depending dc conductivity. 
Furthermore, $\tau_{xx}$ also increases as a function of underlying Cu thickness, but
it is quite different than the time of life of its conducting support.
This is a critical argument in favor of a forming 2DEG interfacial layer:
it will be shown below, in the context of Fig.~\ref{fig:artif-Layer},
that an independent time of life cannot be achieved for an arbitrarily introduced artificial layer.
Also, the non monotonic increment of $\tau_{xx}$ can be associated to
an additional interfacial build-up strain~\cite{sun06prl} possibly induced by 
different densities of oxygen vacancies of \biox\ at the 
interface, as it happens in the STO/LTO 
crystalline or amorphous systems~\cite{Liu2013,li18sr}.

In Table~\ref{tab:param} we also report on the
thickness of the 2DEG layer, $d$. This number provided
minimization of error in the present interface anisotropic model, 
however, this does not
preclude some (fast decaying) extension of the effect further
in $z$-direction as observed in Ref.~[\onlinecite{dubroka10prl}].
Furthermore, since the Bi$_2$O$_3$ layer is already very thick, 20~nm,
the explanation for the non monotonic variation in 2DEG thickness
is not expected to be related to number of Bi$_2$O$_3$ 
monolayers~\cite{park13prb}, but more likely due to the
transition to free-like (for the 20~nm Cu sample) 
from collision-rich (for the 10~nm Cu sample) 
transport along the interface.

\begin{table*}
\caption{\label{tab:param} 
Relevant parameters of the interfacial 2DEG layer
depending on underlying Cu thickness $t_{\mathrm{Cu}}$.
$d$ is the 2DEG thickness, $\sigma^0_{ij}$ and $\tau_{i,j}$
are the DC conductivity and relaxation lifetimes of the 
in-plane and out-of-plane Drude contributions, respectively.
$\omega_i$ 
and $\gamma_i$ ($i=(1,2)$) are SOC$_i$ peaks center and broadening,
respectively.}
\begin{ruledtabular}
\begin{tabular}{c c c c c c c c c c}
  &  & \multicolumn{4}{c}{Free-electron} & \multicolumn{2}{c}{SOC$_1$}   & \multicolumn{2}{c}{SOC$_2$}  \\
\cline{3-6} \cline{7-8} \cline{9-10}  
\rule{0pt}{4ex}$t_{\mathrm{Cu}}$~(nm) & $d$~(nm) & $\sigma^0_{xx}~(10^{3}~$S/cm) & $\tau_{xx}$~(fs)  & 
$\sigma^0_{zz}~$(S/cm) & $\tau_{zz}$~(fs) & $\omega_1$~(meV) & 
$\gamma_1$~(meV) & $\omega_2$~(meV) & $\gamma_2$~(meV) \\
\hline
10 & $0.59\pm0.01$ & $5.0\pm0.01$ & $1.6\pm0.002$ & $18.2\pm0.001$ & $53\pm0.8$ &  $55\pm10$ & $99\pm18$ & -- & -- \\
20 & $0.53\pm 0.02$ & $4.5\pm0.9$ & $9.0\pm1.8$ & $16.0\pm0.3$ & $42\pm0.8$ &  $71\pm4$ & $70\pm3$ & $480\pm6$ & $226\pm3$ \\
50 & $0.82\pm0.005$ & $11.7\pm2.5$ & $18.2\pm3.6$ & $15.8\pm1.2$ & $28\pm2.2$  & $69\pm14$ & $30\pm6$ & $360\pm14$  & $170\pm20$ \\
\end{tabular}
\end{ruledtabular}
\end{table*}

\begin{figure}[thbp]
 \includegraphics[width = 7.5 cm]{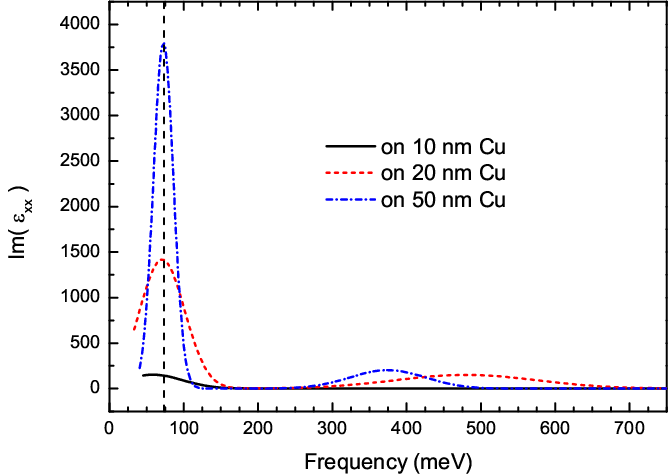}
    \caption{
    (Color online) Imaginary part of the Rashba-enabled spin-orbit coupling (SOC) 
    contributions to the total 2DEG dielectric function shown in Fig.~\ref{fig:sigmaAll}, 
    corresponding to the indicated thickness of the Cu underlayer. 
    The obtained spectral positions,
    widths and amplitudes depend on the 2DEG charge density in fair agreement with 
    theory~\cite{xie14prb}.
    }
\label{fig:soc}
\end{figure}

The rest of the most relevant 2DEG interfacial layer spectral 
features in Fig.~\ref{fig:sigmaAll} are presented next. 
Since the two $\varepsilon_{xx}$ resonances at around 80 and 400~meV, as extracted from 
calculation, 
are consistent with theoretical predictions~\cite{xie14prb} concerning their relative 
amplitudes, frequency positions, and the behavior 
of their frequency-shifts and strengths as a function of 
(Cu-provided) charge density, we will regard them as 
(low frequency) SOC$_1$ and (high-frequency) SOC$_2$, respectively. They are shown
separated from the rest of the contributions in Fig.~\ref{fig:soc}.

The orbital hybridization allowed~\cite{xie14prb} SOC$_1$ resonance
(near the vertical marker in Fig.~\ref{fig:soc}), 
notably increases in strength as function of 
available charge in the interfacial layer, whereas its 
shifting position (to higher frequencies) 
seems to come to a stop for underlying Cu thickness 
above 20~nm. At higher frequencies,
the structure assigned as SOC$_2$, is detected only for the 
thicker Cu samples.
It notably shifts to lower 
frequencies, in opposition to SOC$_1$, and its strength slightly increases with increasing carrier 
density. The behavior of
amplitude ratios, broadening, and energy positions of SOC$_2$ 
and SOC$_1$ in Fig~\ref{fig:soc}, 
are quite consistent with the trend predicted for intersubbands 
optical response as a function of carrier 
density at the 2DEG~\cite{xie14prb}.

The out-of-plane 2DEG dielectric functions $\varepsilon_{zz}$,
in spite of their smallness produce dramatic features in the ellipsometry $\psi$ spectra,
the main one is the already discussed sharp, asymmetric peak at $\sim$75~meV
resulting from the crossing Re $\varepsilon(\omega_{p,z})=0$. As seen in the inset of Fig.~\ref{fig:sigmaAll}(a), the position of this crossing,
which consists of an interplay of particular values $\sigma_{zz}^0$ and $\tau_{zz}$, seems to be 
quite general for the Cu/\biox\ system, whereas the slope is sensitive but only slightly to the
actual properties of the charge density. Moreover, a really similar feature is also seen in the Ag/\biox\ system
(this will be presented elsewhere), which might hint to a stark influence 
of Bi-O orbitals on the properties of the 2DEG~\cite{puebla17apl,Xu2018},
and could be interpreted in the light of a surface-orbital Rashba effect~\cite{go17sr}.
A second feature in $\varepsilon_{zz}$ at $\sim$100$-$200~meV (see insets in 
Fig.~\ref{fig:sigmaAll}),
which was revealed during the iterative procedure described above, was simulated with a 
Tauc-Lorentz oscillator
due to its pulse-like form. This feature in $\varepsilon_{zz}$ is also consistent with theory, 
however, the resolution of
our model only provided a spectrum resembling the envelope of the $z$-conductivity of 
Ref.~[\onlinecite{xie14prb}].
Thermal effects and inherent instrument signal to noise do not permit to resolve subband transitions,
thus we cannot infer whether they are present in the Cu/\biox\ system or not as in the perovskite 
calculations.

Some apparent discrepancies of our results with theory~\cite{xie14prb}, 
besides of the obvious differences between
systems, might be explained by the actual condition of realistic samples. First,
the near percolation threshold conductivity of the 10~nm thick underlying Cu sample
imposes its character to the corresponding 2DEG conductivity. Actually the 2DEG free-electron
contribution exhibits more collisions-related characteristic conductivity than its
underlying Cu. The concomitant spectral weight shift to mid-infrared makes its
Im[$\varepsilon_{xx}$] to look higher than the spectra of 
the other samples [see Fig.~\ref{fig:sigmaAll}(b)], while the 
real parts reflect a more orderly trend.
Also the dc conductivities follow a nice trend (see Table~\ref{tab:param}).
Second, the SOC features in Fig.~\ref{fig:soc} are fairly consistent 
with theory, but show some differences: SOC$_1$ grows with charge 
density but does not shift in frequency when comparing
20 and 50~nm Cu samples, and SOC$_2$, although it presents the 
correct shift, it does not grow noticeably when comparing those same samples. 
This can be explained
by the presence of oxygen vacancies~\cite{torruella17jpcc}, which are critical for the 
acceptance of carriers from the underlying layer in 
perovskite~\cite{Liu2013,li18sr} and \biox\ systems~\cite{Lu18acb}, and might very well be
also relevant in the present Cu/\biox\ system. These vacancies also 
produce a strain field in their vicinity and might influence 
how the orbitals finally hybridize near the interface,
allowing thus for the contrasting properties of \biox\ near the Cu interface
with respect to the \biox\ energy structure in the bulk~\cite{walsh06prb}.

A further utilization of the SiO$_2$/Cu/SiO$_2$ reference samples
consisted on trying to emulate artificially the 
presence of a 2DEG-like intermediate layer in order to
verify whether the Cu/Bi$_2$O$_3$ interface model was real or
an artifact: In Cu/SiO$_2$ the artificial layer dielectric function,
simulated with the isotropic-forced model, and once parameterized with oscillators of the 
underlying Cu and SiO$_2$,
resulted on an effective medium-like combination of
both, depending on the layer placement as shown in the inset of Fig.~\ref{fig:artif-Layer}.
In all the instances described below, the calculated $\psi$ spectra reproduce faithfully the experimental curves.
The artificially introduced layer presents, dominantly, characteristics of a conducting material
of different dc conductivities as seen by the varying heights of the crossing points 
of the Ordal analysis~\cite{ordal83ao}
in Fig.~\ref{fig:artif-Layer}(a), but the frequency of the interceptions, which is directly 
related~\cite{ordal83ao} to the characteristic $\tau$ 
is always preserved and equal to the one of the underlying Cu. Fig.~\ref{fig:artif-Layer}(b) 
shows energy derivatives of the 
real part of the artificial interface's dielectric function, which is also sensitive to changes of 
$\tau$: parallel spectra confirm that
$\tau$ is preserved. The reason for showing derivative spectra plots is that they tend to enhance spectral structures,
and thus show that SiO$_2$ signals are growing when 
fraction of artificial layer depth is more to the
side of SiO$_2$. Furthermore, the fact that the derivative spectra run parallel for the different artificial layer 
placements 
indicates that no other spectral peaks or ``special features"
can be unintentionally obtained by introducing a non-existent layer.
The preservation of characteristic $\tau$ in the artificial layer is in clear contrast to the 
results of Cu/\biox\ interface model, which
produced $\tau_{xx}$ values different from the one of the underlying Cu.

\begin{figure}[htbp]
  \includegraphics[width = 6 cm]{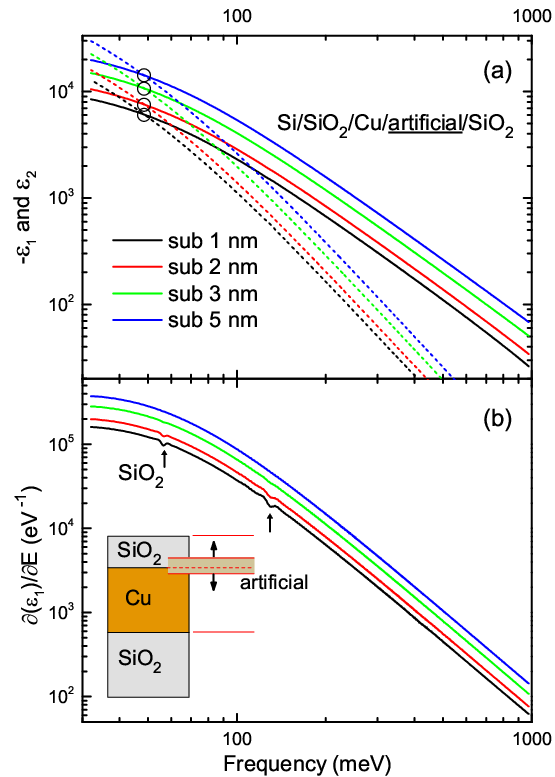}
    \caption{(Color online) Test simulations of a falsely assumed 2DEG-like layer forming
    at the Cu/SiO$_2$ interface for the reference samples following the scheme in the inset. 
    The artificial layer is placed at different depths measured from the actual interface as shown 
    in the scheme.
    The Ordal-type analysis $-\varepsilon_1$ vs $\varepsilon_2$ (solid and dashed lines, respectively)
    in (a) for the thus artificially introduced layer indicates, by the intersections (see the 
    large circles), that the 
    characteristic time of life is the same for all layer depths. This is confirmed by the equal 
    curvatures
    of the energy-derivative of $\varepsilon_1$ in (b).}
\label{fig:artif-Layer}
\end{figure}

\begin{figure}[htbp]
  \includegraphics[width = 6 cm]{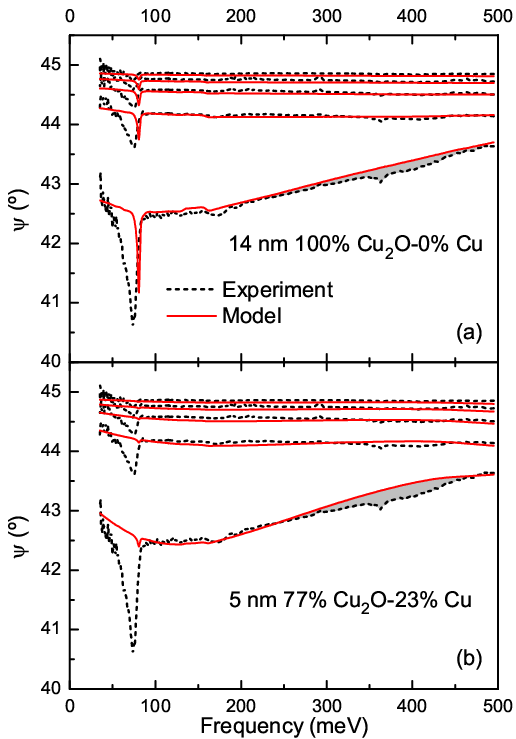}
    \caption{(Color online) Test simulation for the 50~nm Cu/Bi$_2$O$_3$ sample.
    The 2DEG layer is replaced by an effective medium
    comprised of Cu$_2$O and Cu. While the overall $\psi$ values are well reproduced
    for the indicated film thicknesses and proportions,
    obvious discrepancies are noted, specifically, the lack of SOC$_2$ indicated by the
    gray shaded area, and the failure to reproduce the low frequency side of the experiment
    due to the absence of both, SOC$_1$ and the $\varepsilon_{zz}$ crossing effect.} 
\label{fig:ema-CuO-Layer}
\end{figure}

Motivated by the closeness of a large Cu$_2$O IR-active phonon to the sharp experimental feature at $\sim$76~meV,
in Fig.~\ref{fig:ema-CuO-Layer} we present simulations of spectra corresponding to 50~nm Cu/\biox\ 
in which the interfacial layer has been substituted by a medium consisting of a mixture of Cu$_2$O 
and Cu.
The simulations reproduce reasonably well the $\psi$ heights, but
with unrealistic layer thicknesses of the basic constituents which are at odds with our TEM 
measurements
(a separate manuscript is in preparation). The Cu$_2$O$+$Cu layer is also very thick as shown in 
Fig.~\ref{fig:ema-CuO-Layer}(a) and (b). The rationale behind the introduction of Cu in this layer
is to provide it of conducting properties. However higher Cu volume fractions also quenched the 
Cu$_2$O peak.

Other features which cannot be reproduced by these layers include e.g., the SOC$_2$ broad peak, 
which is
highlighted in gray in Fig.~\ref{fig:ema-CuO-Layer}. Although the bulk Cu$_2$O feature has been 
ruled out in our
model, we cannot neglect Cu-O bonding at the interface, which is critical for
the interfacial final orbital hybridization states. The issue here is that its features are either 
out of working range or below our detection limit. In fact, we did not detect any
of the reported~\cite{debbichi12jpcc} Cu oxides by separate confocal Raman studies.
In the case of perovskites, it has been shown that some metals are reactive upon deposition,
in particular the oxidation of Al contributes extensively to the presence of oxygen vacancies
at the SrTiO$_3$ (STO) side, and thus of 2DEG formation\cite{posadas17jap}, which 
shows very large spin-charge interconversion efficiencies~\cite{vaz19natMat}.
It is also known that Cu has poor reactivity to the oxide STO~\cite{posadas17jap}.
From this, we might assume that in the Cu/\biox\ system, Cu will also present poor reactivity,
thus, the dependence of the transport properties with the Cu thickness cannot be explained
by oxidation of the metal, but by the thickness-dependent conducting properties of the support itself.

\begin{figure}[h!tbp]
  \includegraphics[width = 7 cm]{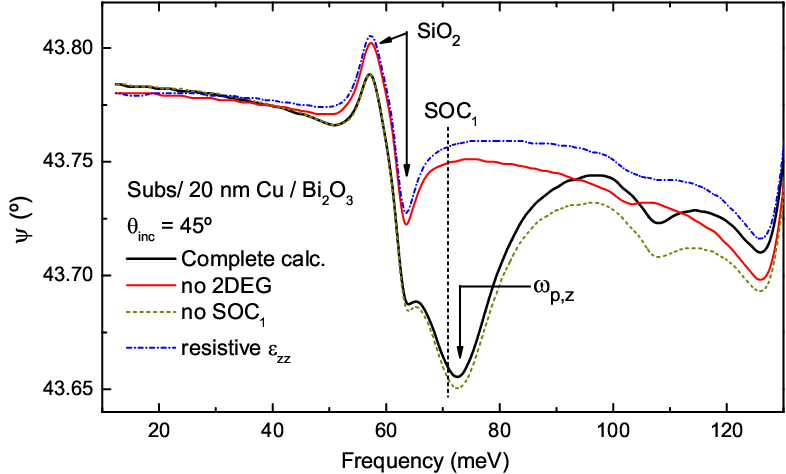}
    \caption{(Color online) Calculated $\psi$ spectra for the 20~nm 
    Cu/Bi$_2$O$_3$ sample for which several features are removed 
    separately, including the complete 2DEG layer, in order to
    observe their influences in the spectral region around 70~meV. 
    }
\label{fig:errorEffects}
\end{figure}

In Fig.~\ref{fig:errorEffects} we present a study of errors induced 
in the calculation by removing some features
whose frequencies are very close together in the range of 10 to 130~meV, 
for which analytic oscillator functions (in the sense of ``in opposition to numerical data'') where continued
below the experimental range (initial frequencies are $30-40$~meV, 
depending on light availability) in order to discard the 
influence of undesired features out of the working spectral range.
We compare the final result of Fig.~\ref{fig:fitExp}(b) for the 
$45^\circ$ angle of incidence, which has a significant projection 
of $x$ component of incident $p$ electric field, to several 
situations.

The complete removal of the 2DEG layer, which also induces an error 
due to film thickness, provides an idea of
our starting point. This spectrum reveals that the Cu dielectric function 
and thickness, as extracted from reference samples, 
are appropriate for the Cu/\biox\ system, since the SiO$_2$
features in this range seem fairly correct
concerning size and phase of the feature. 
The influence of the conducting part of the $z$-component is seen 
by simulating the 2DEG layer with a $\varepsilon_{zz}$ of 
resistive character, which therefore lacks the 
$\omega_{p,z}$ crossing.
We conclude that its presence is the most significant contribution 
for the $\sim$72~meV negative peak in $\psi$.
The subbands feature of $\varepsilon_{zz}$ also helps to correctly place 
a small kink at $\sim$110~meV.
Its presence in experiment is more clearly seen in the 
$\theta_\mathrm{inc}=85^\circ$ spectrum in 
Fig.~\ref{fig:fitExp}(b).
In this series of simulations we also removed the oscillator 
corresponding to SOC$_1$. 
In Fig.~\ref{fig:errorEffects} we have indicated the center of this 
resonance with the vertical line. Calculations show that
the real part of the oscillator for this layer has more influence in 
$\psi$, whereas its peak-like imaginary part contributes more
to refinement of $\Delta$ spectra. 
Fig.~\ref{fig:errorEffects} probes that SOC$_1$ has a more 
delocalized impact on the spectrum, which is
more noticeable at higher frequencies from its center. 
In Table~\ref{tab:param}, some of the parameters show large errors, sometimes of
the order of $20\%$. This can be attributed to high correlations among different parameters
and thus a great difficulty to clearly separate contributions, in particular those present at the
low frequency side of the working range. This can be illustrated, for example, by comparing
in-plane conductivity parameters: the Drude characteristics $\sigma_{xx}^0$ 
and $\tau_{xx}$ for the 20~nm and 50~nm Cu samples
present large errors partly due to the featureless character of the classical Drude line shape
which can be contrasted to the 10~nm Cu sample resulting parameters, 
which show nearly negligible uncertainties, possibly since
its percolation-related  mid infrared peak allows for an easier deconvolution. 
In the case of the SOC$_1$,
the difficulty to be extracted and its concomitant error are very likely consequence of its
non-local influence on the overall spectrum as mentioned above, in Fig.~\ref{fig:errorEffects}.

\section{Conclusions}
In summary, we have detected the presence of a 2DEG
at the interface of an amorphous oxide with a non magnetic metal
by means of infrared ellipsometry.
We have found that a diagonal, uniaxial dielectric tensor of the
2DEG layer describes really well the experimental results.
Furthermore, beside the substrate dependent carrier density free-electron (Drude-like) 
behavior, we have also observed
intersubband features consistent with SOC of the Rashba type.
The present report sheds light on the complex phenomena at amorphous interfaces that have been elusive for decades.
This work allows, therefore, assessing the complex phenomena associated to the presence of 2DEG with SOC at interfaces between 
amorphous materials. Beyond the particular case of study in the present paper, the supporting theoretical model 
by other groups and the advanced ellipsometry technique suggests applicability in a larger range of unexplored amorphous interfaces.

\begin{acknowledgments}
We thank Prof. A. H. MacDonald (University of Texas at Austin) and 
Prof. J. J. Palacios (Universidad Aut\'onoma de Madrid) for insightful guidance. 
The expert technical support of Esequiel Ontiveros, F. Ram{\'\i}rez-Jacobo, and  
L.E. Guevara-Mac{\'\i}as is highly appreciated. 
We also thank Consejo Nacional de Ciencia y Tecnolog{\'\i}a (Mexico) for Grants CB-223564-2014, 
CB-252867-2015, 299552 and 206298 (IRSE equipment). 
This work was also supported by a Grant-in-Aid for Scientific Research on Innovative Area, 
``Nano Spin Conversion Science" (Grant No. 26103002) and RIKEN incentive Research Project Grant 
No. FY2016.
\end{acknowledgments}

\appendix*
\section{Fresnel coefficients}

For clarity of how the anisotropic layer is handled in the model, we reproduce here the main 
equations employed to calculate the Fresnel coefficients 
(see e.g., Ref.~[\onlinecite{FujiwaraSEllipsom}]).
The reflection coefficients at the interfaces of the 2DEG layer,
which according to the layers numbered in Sec.~\ref{sec:model} and Fig.~\ref{fig:Scheme} 
is layer number (2), for the case of tensor orientation
$\varepsilon_{xx}=\varepsilon_{yy}\neq \varepsilon_{zz}$ are
\begin{eqnarray}
r_{12}^\mathrm{p} & = & \frac{n_{2,x} n_{2,z}\cos\theta_1-n_{1}
	\left(\varepsilon_{2,z}-\varepsilon_1\sin^2\theta_1\right)^{1/2}}{
	n_{2,x} n_{2,z}\cos\theta_1+n_{1}
	\left(\varepsilon_{2,z}-\varepsilon_1\sin^2\theta_1\right)^{1/2}}, \\
r_{12}^\mathrm{s} & = & \frac{n_{1}\cos\theta_1-
	\left(\varepsilon_{2,x}-\varepsilon_1\sin^2\theta_1\right)^{1/2}}{
	n_{1}\cos\theta_1+
	\left(\varepsilon_{2,x}-\varepsilon_1\sin^2\theta_1\right)^{1/2}
	}, 	
\end{eqnarray}

\noindent for the \biox/2DEG interface, where the complex refractive index $n=\sqrt{\varepsilon}$ for nonmagnetic materials.
The terms in brackets correspond to the angle of propagation within the 2DEG layer expressed in terms of
the propagation angle within the \biox\ layer immediately above. These expressions are straightforwardly reduced to the isotropic case~\cite{FujiwaraSEllipsom}, so they can be used for other interfaces. For the 2DEG/Cu interface the reflection coefficients are

\begin{eqnarray}
r_{23}^\mathrm{p} & = & \frac{n_3 \left(\varepsilon_{2,z}-\varepsilon_3\sin^2\theta_3\right)^{1/2}-n_{2,x}n_{2,z}\cos\theta_2}
	{n_3 \left(\varepsilon_{2,z}-\varepsilon_3\sin^2\theta_3\right)^{1/2}  +  n_{2,x}n_{2,z}\cos\theta_2}, \\
r_{23}^\mathrm{s} & = & \frac{\left(\varepsilon_{2,x}-\varepsilon_3\sin^2\theta_3\right)^{1/2}-n_3\cos\theta_3}
	{\left(\varepsilon_{2,x} - \varepsilon_3\sin^2\theta_3\right)^{1/2}  +  n_3\cos\theta_3}. 	
\end{eqnarray}

As the polarized light propagates through the anisotropic layer, the phase of each of its components is affected differently, and this is accounted for by using the standard model of multiple reflections through

\begin{eqnarray}
\beta_2^\mathrm{p} & = & \frac{2\pi d}{\lambda} \frac{n_{2,x}}{n_{2,z}} \left(\varepsilon_{2,z}-\varepsilon_1\sin^2\theta_1\right)^{1/2}\\
\beta_2^\mathrm{s} & = & \frac{2\pi d}{\lambda}  \left(\varepsilon_{2,x}-\varepsilon_1\sin^2\theta_1\right)^{1/2}
\end{eqnarray}
\noindent where $\lambda$ is the wavelength of the incident radiation. For the isotropic layers the $\beta$ phases are also isotropic.  

The Fresnel coefficients are calculated for each polarization by introducing the interfacial reflection coefficients at both sides of the $b$-th layer, say, using the previously calculated reflection coefficients 
$r_{ab}^i$ and $r_{bc}^i$, and the intra-layer phase shifts $\beta_b^i$, where $i=(s,p)$, by moving from bottom to top of the stacked structure~\cite{osheavens}

\begin{equation}
\label{eq:Fresnel}
r_{ab}, r_{bc}, \beta_b \rightarrow r_{abc} = \frac{r_{ab}+r_{bc}e^{-i 2 \beta_b}}{1+r_{ab} r_{bc}e^{-i 2 \beta_b}}.
\end{equation}
\noindent The Fresnel coefficients are calculated in a cumulative fashion: starting with Si and SiO$_2$, i.e., layers 4 and 5 according to the stacked structure presented in Sec.~\ref{sec:mod:mater} [see also Fig.~\ref{fig:Scheme}(a)], the Fresnel coefficients $r_{345}^i$ are calculated, then all indexes are decremented and the Fresnel coefficient  $r_{abc}$ takes the place of the previous interfacial reflection coefficient $r_{bc}$, i.e, $r_{2345}$ is calculated with $r_{23}$ and $r_{345}$. The process is
repeated until the Fresnel coefficient of the whole stack $r_{01...5}$ is finally calculated. 
This has to be done separately for both $i = (p,s)$ polarizations. Thus, the quantities to extract, dielectric functions and thicknesses, are included, correspondingly, within polarization dependent reflection coefficients and $\beta$ exponents. 



\bibliography{v1Bi2O3-bib}
\bibliographystyle{apsrev}

\end{document}